\documentclass{cdclass}
\usepackage{amsmath,amsfonts}
\usepackage{ulem}
\usepackage{xcolor}

\newcommand{\ce}[1]{{\color{black} #1}}
\newcommand{\dece}[1]{{\color{black} #1}}

\renewcommand{\sout}[1]{\unskip}

\begin{document}
	
	\title{Mean-Field Games Modeling of Anticipation in Dense Crowds}
	
	\author{
		Matteo Butano\authorlabel{1} \and 
		Cécile Appert-Rolland\authorlabel{2} \and
		Denis Ullmo\authorlabel{3}
	}
	\authorrunning{M. Butano \and C. Appert-Rolland \and D. Ullmo}
	\institute{
		\authorlabel{1} CNRS, Université Paris-Saclay, LPTMS, IJCLab
		\authoremail{1}{matteo.butano@universite-paris-saclay.fr}
		\and
		\authorlabel{2} CNRS, Université Paris-Saclay, IJCLab, 
		\authoremail{2}{Cecile.Appert-Rolland@ijclab.in2p3.fr}
		\and
		\authorlabel{3} CNRS, Université Paris-Saclay, LPTMS
		\authoremail{3}{denis.ullmo@universite-paris-saclay.fr}
	}
	
	\date{year}{date1}{date2}{date3} 
	\ldoi{10.17815/CD.20XX.X} 
	\volume{V}  
	\online{AX} 
	
	\maketitle
	
	\nolinenumbers
	\begin{abstract}
		Understanding and modeling pedestrian dynamics in dense crowds is a complex yet essential aspect of crowd management and urban planning. In this work, we investigate the dynamics of a dense crowd crossed by a cylindrical intruder using a Mean-Field Game (MFG) model. By incorporating a discount factor to account for pedestrians' limited anticipation and information processing, we examine the model's ability to simulate two distinct experimental configurations: pedestrians facing the obstacle and pedestrians giving their back to the intruder. Through a comprehensive comparison with experimental data, we demonstrate that the MFG model effectively captures essential crowd behaviors, including anticipatory motion and collision avoidance. 
	\end{abstract}
	
	
	\section{Introduction}
	\label{sec:intro}
	
	Understanding and modeling pedestrian motion in dense crowds is a challenging task due to its multi-scale nature. Traditionally \cite{hoogendoorn2004pedestrian}, pedestrians behavior has been categorized into strategic, tactical, and operational levels, each addressing different aspects of their movement. The strategic level encompasses travel goals and general timing, while the tactical level involves route selection. At the operational level, the focus shifts to the actual execution of movement on the chosen route. While long-term optimization naturally applies to strategic and tactical levels, capturing the operational level, especially in dense crowd scenarios, is \dece{often} done considering dynamical models incorporating physical and social forces \cite{helbing1995social}. However, \dece{the comparison~\cite{Bonnemain_Butano_ped_not_grains_players} with experimental data of \cite{Nicolas_Kuperman_mech_response}  involving the crossing of a dense crowd by a cylindrical intruder, showed that such models fail when \dece{ the anticipation takes place on a time scale significantly longer than the time to the next collision}. By contrast, comparison with the same experimental data} highlighted how a model based on Mean-Field Games can accurately describe the long-term anticipation and competitive optimization mechanisms at play in such scenario~\cite{Bonnemain_Butano_ped_not_grains_players}.
	
	In this paper we discuss how expanding the MFG model, adding a term discounting future information, referred to as the ``discount factor'', better captures, in certain situations, the behavior of a dense crowd crossed by a cylindrical intruder. The discount factor serves as a term that attenuates the significance of future events beyond a certain cutoff, effectively modeling the pedestrians' shortsightedness and their limited access to perfect information. Our objective is to investigate if this improvement is enough to reproduce the experimental configurations of \cite{Nicolas_Kuperman_mech_response} that were not treated in \cite{Bonnemain_Butano_ped_not_grains_players}. We aim to demonstrate that the introduction of the discount factor can effectively account for situations where pedestrians have limited knowledge or are unable to anticipate events too far into the future.
	
	In the following sections, we will present the mean-field game model enriched with the discount factor and discuss its theoretical implications. We will then compare the model's predictions with experimental data obtained from different configurations involving a static dense crowd crossed by a cylindrical intruder. By examining these scenarios, we will analyze how the discount factor improves the model's accuracy in capturing the observed crowd behavior in the presence of limited information and anticipation.
	
	\section{The experimental results}
	\label{sec:experiment}
	
	The experiment described in \cite{Nicolas_Kuperman_mech_response,appert-rolland2020b} aimed to investigate the behavior of pedestrians in a dense crowd crossed by a cylindrical intruder. \dece{Experimental data consist in sets of individual trajectories. In order to compare with the MFG model of the next section, we need to build a density field from these trajectories. There are several ways to perform this task. In~\cite{Nicolas_Kuperman_mech_response}, a reconstruction based on Voronoi cells was used. This gives a density field which is by construction flat near and up to the cylinder. Here, as the MFG approach describes the depletion on the frontier of the cylinder, we chose a method preserving this property, and the density field was obtained from a Gaussian convolution.}The experiment was performed with people arranged in \dece{various configurations. In the following, we shall consider two of them}: (a) facing the obstacle, \dece{and}  (b) giving their back to the cylinder and being told not to anticipate.
    \begin{figure}
		\centering
		\includegraphics[width=0.7\linewidth]{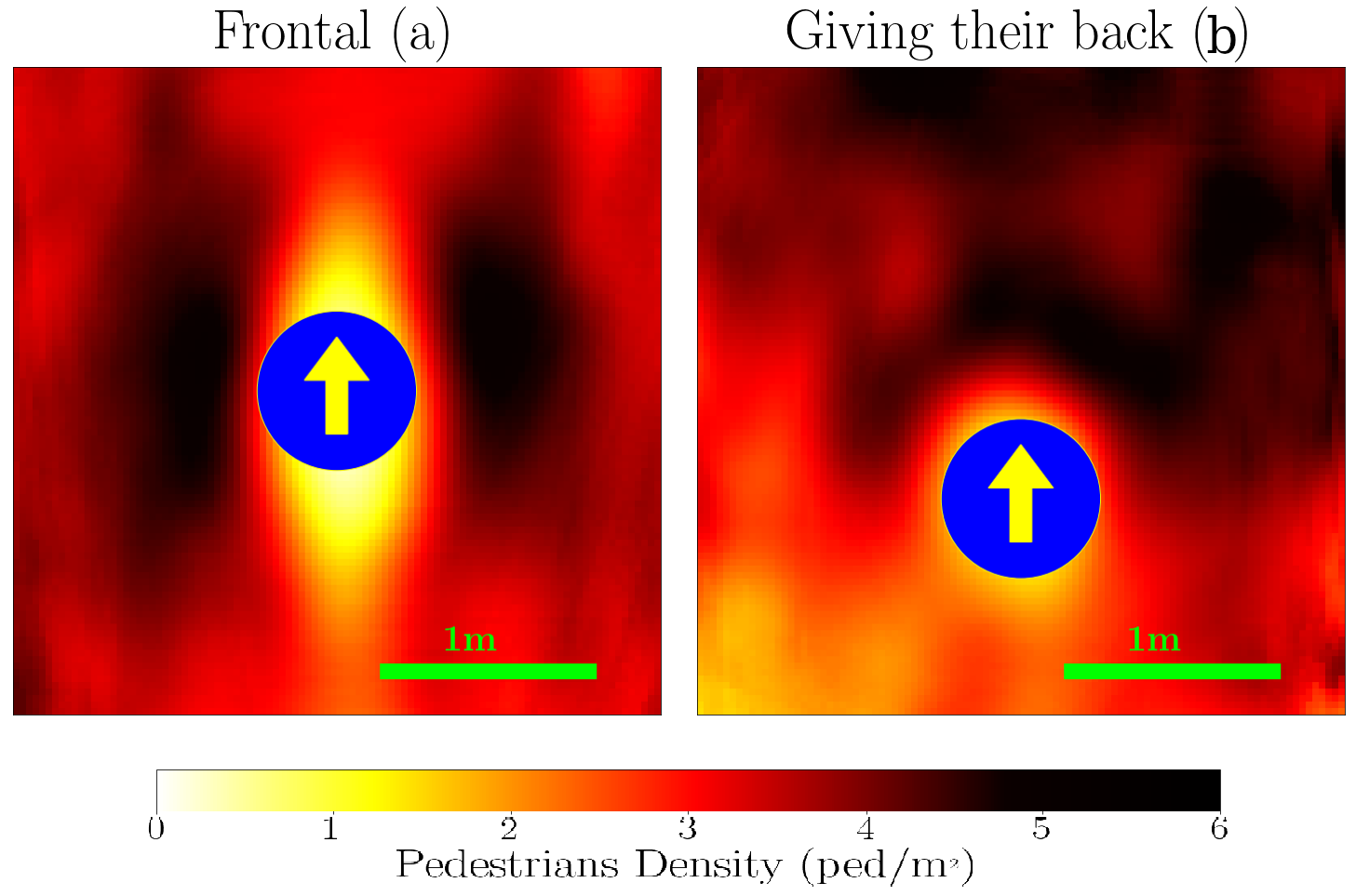}
		\caption{Density plots of the passage of a cylindrical intruder through a dense crowd from the experimental data collected in \cite{Nicolas_Kuperman_mech_response} in \dece{two} different configurations: \dece{(left)} participants facing the intruder, \dece{(right) participants} giving their back to it while actively not anticipating.}
		\label{fig:1}
	\end{figure}
	\begin{itemize}
		\item[(a)]\textbf{Facing the Obstacle}: In this configuration, participants faced the approaching cylinder. As the obstacle advanced, pedestrians displayed anticipatory behavior by temporarily accepting higher local densities to expedite the obstacle's removal. They moved outward to avoid the intruder's arrival and inward to regain less congested positions, \dece{with displacements} always oriented perpendicularly to the motion of the obstacle. This anticipatory dynamics demonstrates that participants possessed perfect information about the system and were able to plan their actions accordingly.
		\item[(b)] 
  \textbf{Giving Their Back to the Obstacle} In this case, participants were asked to give their back to the approaching intruder, effectively looking away from it, and, since they were still able to sense the obstacle's arrival, they were asked not to anticipate. This setup eliminated any direct information about the obstacle's approach. Interestingly, the observed behavior shifted dramatically with the accumulation patterns moving from the sides to the front of the obstacle, mimicking situations found in granular matter systems. 
	\end{itemize}
		
	By introducing the discount factor, we aim to enhance the MFG model's ability to capture the observed dynamics in these configuration, thereby shedding light on the influence of information and anticipation on the collective behavior of pedestrians during intruder crossings. The following section will introduce the MFG model and discuss its theoretical implications, followed by a comparative analysis of the model's predictions with the experimental data from the three configurations, further validating its effectiveness in capturing the complex behavior of dense crowds.
	
	\section{The Mean-Field Games model}
	\label{sec:mfg_model}
	
	Mean-Field Games (MFG) constitutes a relatively new field of research. Its foundations are in the works of J.-M. Lasry and P.-L. Lions \cite{Lasry_Lions_horizon_fini,Lasry_Lions_stationnaire}, and of M. Huang, R. P. Malhamé and P. E. Caines  \cite{Huang_Malhamé_large_pop_stoch_dyn}. During the years, many works have been focused on the mathematical properties of MFG \cite{Cardaliaguet_notes_mfg,Gomes_Saude_mfg_survey,Bensoussan_Frehse_master_eq_mft,Cardaliaguet_Delarue_master_eq_convergence_prob,Carmona_Delarue_prob_analysis_mfg}. Although there are applications of MFG to pedestrian dynamics \cite{Lachapelle_Wolfram_mfg_cong}, to the best of our knowledge comparisons of crowds simulated with MFG to experimental data \cite{Bonnemain_Butano_ped_not_grains_players} are rare. A general and mathematically rigorous discussion of the foundations of MFG being found in the book \cite{Cardaliaguet_Delarue_master_eq_convergence_prob}, a physicist-friendly version being  exposed in \cite{Ullmo_Swiecicki_quadratic_mfg}, here we will limit ourselves to essentials.
	
	\subsection{The discounted equations}
	
	In the specific settings of our MFG model, each agent's \textit{state variable} $ \vec{X}(t) \in \mathbb{R}^2$, representing their position, evolves following the \textit{Langevin equation} 
	\begin{equation}
		\label{eqn:langevin}
		\dot{\vec{X}} = \vec{a}(t)+\sigma\vec{\xi}(t),
	\end{equation}
	where $\vec{\xi}(t)$ is a two dimensional Gaussian white noise, and $\vec{a}$ is the drift velocity,  the \textit{control parameter} that represents the strategy players choose by minimizing the discounted cost functional defined, in this case, as 
	\small
	\begin{equation}
		\label{def:cost_disc}
		c[\vec{a}](\vec{x},t)=\mathbb{E}\left\{\int_t^T \mathcal{L}(\vec{x},\tau)[m] e^{\gamma(t-\tau)}d\tau + e^{\gamma(t-T)}c_T(\vec{x}_T) \right\}
	\end{equation}
	\normalsize
	where $ c_T $ is a \textit{terminal cost}, that could be used to represent a target for pedestrians, e.g. an exit,  $\gamma$ is the discount factor, that tells how far into the future agents will look while optimizing, and
	\begin{equation}
		\label{def:lagrangian}
		\mathcal{L}(\vec{x},\tau)[m] = \frac{\mu}{2}(\vec{a}(\tau))^2 - V[m](\vec{x},\tau)
	\end{equation}
	can be seen as the term describing the agents' preferences. In fact, the squared velocity tells that going too fast is detrimental, and that the best would be to stand still, but the presence of the external world, represented by the potential term
	\begin{equation}
		\label{def:potential}
		V[m](\vec{x},t) = gm(\vec{x},t) + U_0(\vec{x},t),
	\end{equation}
	describing the interaction with the others and with the environment, cause agents to actually move. Here the environment, \dece{described by the term  $U_0(\vec{x},t) $,} would be the moving cylinder, but different scenarios could be accounted for. The main assumptions of MFG are that all agents are equivalent and the interaction with others is of mean-field type, determined only through the average density 
	\begin{equation}
		m(\vec{x},t) =\lim_{N\rightarrow +\infty}\mathbb{E}\left[ \frac{1}{N}\sum_{i = 1}^{N}\delta(\vec{x}-\vec{X}_i(t))\right] \; .
	\end{equation}
The quantity of interest is then the \textit{value function}
	\begin{equation}
		u(\vec{x},t) = \inf_{\vec{a}}c[\vec{a}](\vec{x},t).
	\end{equation}
	At this point, using Ito's calculus and the \textit{dynamic programming principle}, we find that the value function solves the following \textit{Hamilton-Jacobi-Bellman} equation
	\begin{equation}
		\label{eqn:HJB_disc}
		\begin{cases}
			\partial_t u  = -\frac{\sigma^2}{2}\Delta{u} + \frac{1}{2\mu}(\vec{\nabla}u)^2 + \gamma u +  V[m] \\
			u(\vec{x},t = T) = c_T(\vec{x})
		\end{cases}
	\end{equation}
	This is a backward equation, solved starting from the terminal condition $ c_T(\vec{x}) $, which is useful to set goals to be reached by time $ T $ by the agents, e.g. exiting a room or reaching a certain area. Then, given the stochastic evolution of each player's state variable, we know that the corresponding density evolves following the Kolmogorov-Fokker-Plank equation 
	\begin{equation}
		\label{eqn:KFP_disc}
		\begin{cases}
			\partial_t m = \frac{\sigma^2}{2}\Delta m  +\frac{1}{\mu}\nabla\cdot(m\nabla u)\\
			m(\vec{x},t = 0) = m_0(\vec{x})
		\end{cases}
	\end{equation}
	a forward equation solved starting from an initial density profile.
	
	\section{Results}
	\label{sec:results}
	
	\begin{figure}
		\centering
		\includegraphics[width=0.6\linewidth]{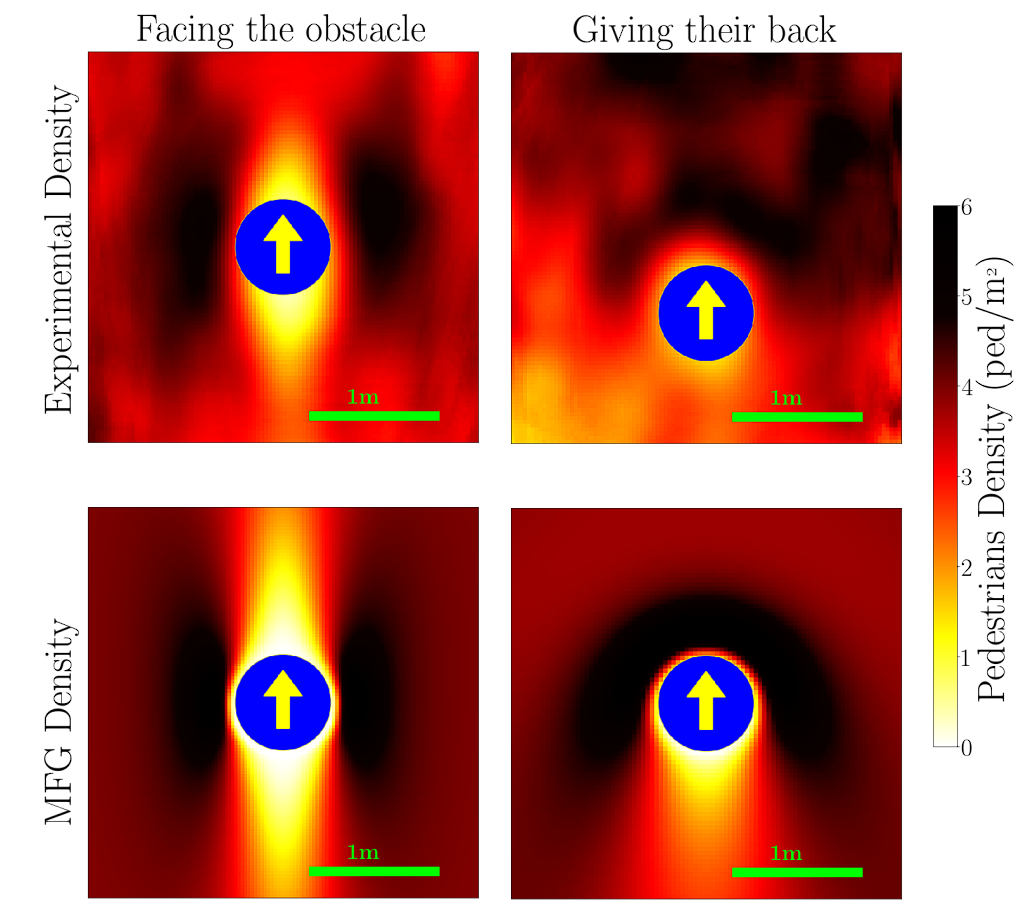}
		\caption{Comparison between the experimental data (top row) of the density of a crowd crossed by a cylindrical intruder \cite{Nicolas_Kuperman_mech_response} in \dece{two} different configurations (\dece{\sout{from left to right}left:} participants facing the intruder, \dece{right:} giving their back to it while actively not anticipating), and the MFG simulation \dece{(bottom row)} with different values of the discount factor (\dece{left:} $\gamma = 0Hz$,  \dece{right:} $\gamma = 6Hz$\dece{).\sout{respectively from left to right)}}}
		\label{key}
	\end{figure}
	
	We applied our Mean-Field Games model to the case where pedestrians were facing the obstacle, as previously studied in \cite{Bonnemain_Butano_ped_not_grains_players}. Since in this configuration participants have complete information, we chose to use $\gamma = 0Hz$, \dece{corresponding to an infinite anticipation time $1/\gamma$.} With this settings we accurately replicated the vertically symmetric distribution of pedestrians, with density depletion prior and posterior to the obstacle passage and an increase on the sides. \dece{As shown in \cite{Bonnemain_Butano_ped_not_grains_players}}, in this case the model also correctly displays the lateral motion of pedestrians stepping aside to accommodate the intruder, effectively capturing the long-term anticipatory behavior present in the experiment.

	Next, 
 we focused on the scenario where pedestrians gave their back to the obstacle and were asked not to anticipate. The observed behavior in this case differed significantly, with pedestrians being pushed along by the intruder, similar to granular material. The MFG simulation with $\gamma = 6Hz$ and the other parameters appropriately adjusted to fit the data, recovered the accumulation in front of the obstacle and the smaller depletion behind it. The velocity plot, as it will be shown in more detailed publication \cite{Butano2023}, correctly displays pedestrians both being pushed along by the intruder and rotating around it to escape the congestion. 
	
	\section{Conclusion}
	\label{sec:conclusion}
	
	In conclusion, our study explores the application of an enriched Mean-Field Game (MFG) model, incorporating a discount factor to account for pedestrians' limited anticipation and information processing, to the simulation of a dense crowd crossed by a cylindrical intruder. Through a comparison with experimental data from two distinct configurations, we have shown that the MFG model effectively captures essential crowd behaviors, such as anticipatory motion and collision avoidance. The introduction of a discount factor allows the model to adapt to various scenarios, demonstrating its versatility. \dece{More properties of the model will be detailed further in future publication~\cite{Butano2023}.}
 
 These results 
 shed light on the importance of long-term anticipation in crowd behavior\dece{, a feature that would deserve further investigation, in particular through experiments or data analysis. Indeed, it may play a role in the arising or prevention of stampedes or of unwanted propagating waves in dense crowds.}

	\begin{acknowledgements}
		\dece{We thank Alexandre Nicolas, Marcelo Kuperman, Santiago Ibañez, and Sebastián Bouzat, for the experimental data on which the comparison is based.} We also thank the CNRS for providing the funding to Matteo Butano's PhD through the Imperial College-CNRS Joint PhD program.
        \ce{\sout{MATTEO, IL FAUT SANS DOUTE NOMMER LE FINANCEUR DE TA THESE?}}
	\end{acknowledgements}
	
	
	\begin{contributions}
		\dece{Cécile-Appert Rolland participated to the organization of the experiments of \cite{Nicolas_Kuperman_mech_response}. All authors worked on the development of the $\gamma = 0$ version of the model, that Denis Ullmo and Matteo Butano extended to the $\gamma \neq 0$ case.} 
	\end{contributions}
	
	\newpage
	
	\bibliographystyle{cdbibstyle} 
	\bibliography{MFG_biblio} 
\end{document}